
\input harvmac.tex
\input epsf.tex
\def\emrvac{|\emptyset \rangle }
\def\emlvac{\langle \emptyset |}
\def\phih{\hat{\phi}}
\def\alt{\tilde{\al}}

\def\mass{{\rm {\hat{m}} }}
\def\ma{\mass}
\def\ha{ {\scriptstyle{\inv{2}} }}
\def\mtoo{{\mathop{\longrightarrow}^{\ma \to 0}}}

\def\betah{{\hat{\beta}}}

\def\hf{\CH_F}

\def\slh{{\hat{sl(2)}}}
\def\psib{\bar{\psi}}
\def\psip{\psi_+}
\def\psim{\psi_-}
\def\psibp{\psib_+}
\def\psibm{\psib_-}

\def\vphi{\varphi}
\def\bh{\hat{b}}

\def\sqm{\sqrt{\ma}}
\def\ez{e^{\ma z u + \ma \zb /u}}
\def\squ{\sqrt{u}}
\def\bp{b^+}
\def\bm{b^-}
\def\bb{\bar{b}}
\def\bbp{\bar{b}^+}
\def\bbm{\bar{b}^-}
\def\du{ \frac{du}{2 \pi i |u|} }
\def\dua{ \frac{du}{2 \pi i u} }

\def\Qpm#1{ Q^\pm_#1 }

\def\mn{\frac{ \ma^{|n|+|m|} }{\ma^{|n+m|} } }

\def\slhh{\hat{\hat{sl(2)}}}

\def\lvacpm{\langle \pm \ha \vert}
\def\rvacpm{\vert \pm \ha \rangle}
\def\rvacmp{\vert \mp \ha \rangle}
\def\rvacp{\vert + \ha \rangle}
\def\rvacm{\vert - \ha \rangle}
\def\va#1{\vert {#1} \rangle}
\def\lva#1{\langle {#1} \vert}

\def\col#1#2{\left(\matrix{#1\cr #2\cr}\right)}
\def\dz{\frac{dz}{2\pi i}}
\def\dzb{\frac{d\zb}{2\pi i}}

\def\uone{U(1)}
\def\psipm{\psi_\pm}
\def\psibpm{\psib_\pm }
\def\bhpm{\bh^\pm}
\def\bpm{b^\pm}
\def\bbpm{\bar{b}^\pm}

\def\bhpm{\bh^\pm}
\def\cL{\CC^{L}_{\vphi}}
\def\cR{\CC^{R}_{\vphi} }

\def\hal{{\CH^L_a}}

\def\vhh{\hat{\hat{V}}}

\def\listrefs{\footatend\immediate\closeout\rfile\writestoppt
\baselineskip=14pt\centerline{{\bf References}}\bigskip{\frenchspacing%
\parindent=20pt\escapechar=` \input \jobname.refs\vfill\eject}\nonfrenchspacing}

%
%
%
%

\def\tilde{\widetilde}
\def\bar{\overline}
\def\hat{\widehat}
\def\*{\star}
\def\[{\left[}
\def\]{\right]}
\def\({\left(}		
\def\){\right)}

%
%
\def\zb{{\bar{z} }}
\def\frac#1#2{{#1 \over #2}}
\def\inv#1{{1 \over #1}}

\def\d{\partial}

\def\rvac{\hbox{$\vert 0\rangle$}}
\def\lvac{\hbox{$\langle 0 \vert $}}
\def\2pi{\hbox{$2\pi i$}}

\def\dsl{\raise.15ex\hbox{/}\kern-.57em\partial}
\def\Dsl{\,\raise.15ex\hbox{/}\mkern-.13.5mu D}
%
%
\def\th{\theta}		
		\def\Ga{\Gamma}

\def\al{\alpha}
\def\ep{\epsilon}
\def\la{\lambda}	\def\La{\Lambda}
\def\de{\delta}		
\def\om{\omega}		
	
\def\vphi{\varphi}
%
%
		\def\CC{{\cal C}}
		\def\CF{{\cal F}}
	\def\CH{{\cal H}}

\def\rvac{\hbox{$\vert 0\rangle$}}
\def\lvac{\hbox{$\langle 0 \vert $}}

\def\2pi{\hbox{$2\pi i$}}

\def\dsl{\raise.15ex\hbox{/}\kern-.57em\partial}
\def\Dsl{\,\raise.15ex\hbox{/}\mkern-.13.5mu D}
%
%
%
\font\numbers=cmss12
\font\upright=cmu10 scaled\magstep1
\def\stroke{\vrule height8pt width0.4pt depth-0.1pt}
\def\topfleck{\vrule height8pt width0.5pt depth-5.9pt}
\def\botfleck{\vrule height2pt width0.5pt depth0.1pt}
\def\Zmath{\vcenter{\hbox{\numbers\rlap{\rlap{Z}\kern
0.8pt\topfleck}\kern
2.2pt
                   \rlap Z\kern 6pt\botfleck\kern 1pt}}}
\def\Qmath{\vcenter{\hbox{\upright\rlap{\rlap{Q}\kern
                   3.8pt\stroke}\phantom{Q}}}}
\def\Nmath{\vcenter{\hbox{\upright\rlap{I}\kern 1.7pt N}}}
\def\Cmath{\vcenter{\hbox{\upright\rlap{\rlap{C}\kern
                   3.8pt\stroke}\phantom{C}}}}
\def\Rmath{\vcenter{\hbox{\upright\rlap{I}\kern 1.7pt R}}}
\def\Z{\ifmmode\Zmath\else$\Zmath$\fi}
\def\Q{\ifmmode\Qmath\else$\Qmath$\fi}
\def\N{\ifmmode\Nmath\else$\Nmath$\fi}
\def\C{\ifmmode\Cmath\else$\Cmath$\fi}
\def\R{\ifmmode\Rmath\else$\Rmath$\fi}
\def\Zmath{Z}


\def\AASS{E. Abdalla, M. C. B. Abdalla, G. Sotkov, and M. Stanishkov,
{\it Off Critical Current Algebras}, Univ. Sao Paulo preprint,
IFUSP-preprint-1027, Jan. 1993.}

\def\Griffin{P. Griffin, Nucl. Phys. B334, 637.}
\def\MSS{B. Schroer and T. T. Truong, Nucl. Phys. B144 (1978) 80 \semi
E. C. Marino, B. Schroer, and J. A. Swieca, Nucl. Phys. B200 (1982) 473.}

\def\BLnlc{D. Bernard and A. LeClair, Commun. Math. Phys. 142 (1991) 99;
Phys. Lett. B247 (1990) 309.}
\def\form{F. A. Smirnov, {\it Form Factors in Completely Integrable
Models of Quantum Field Theory}, in {\it Advanced Series in Mathematical
Physics} 14, World Scientific, 1992.}

\def\ZZ{A. B. Zamolodchikov and Al. B. Zamolodchikov, Annals
Phys. 120 (1979) 253.}
\def\Colemani{S. Coleman, Phys. Rev. D 11 (1975) 2088.}
\def\TI{H. Itoyama and H. B. Thacker, Nucl. Phys. B320 (1989) 541.}
%
%
%

%

\Title{CLNS 93/1255}
{\vbox{\centerline{ Affine Lie Algebras in Massive Field Theory   }
\centerline{ and } \centerline{Form Factors from Vertex Operators} }}

\bigskip
\bigskip

\centerline{Andr\'e LeClair}
\medskip\centerline{Newman Laboratory}
\centerline{Cornell University}
\centerline{Ithaca, NY  14853}
\bigskip\bigskip

\vskip .3in

We present a new application of affine Lie algebras to
massive quantum field theory in 2 dimensions, by investigating
the $q\to 1$ limit of the q-deformed affine $\hat{sl(2)}$  symmetry of
the sine-Gordon theory, this limit occurring at the free fermion point.
Working in radial quantization leads to a quasi-chiral factorization of
the space of fields.
The
conserved charges which generate the
affine Lie algebra split into two independent affine algebras on
this factorized space, each with level 1 in the anti-periodic sector,
and level $0$ in the periodic sector.
The space of fields in the anti-periodic sector can be organized
using level-$1$ highest weight representations, if one supplements
the $\slh$ algebra with the usual local integrals of motion.
Introducing a particle-field duality leads to a new way of computing
form-factors in radial quantization.
Using the integrals of motion, a momentum
space bosonization involving vertex operators is formulated.
Form-factors are computed as
vacuum expectation values of vertex operators in momentum space.
{\it Based on talk given at the Berkeley Strings 93 conference,
May 1993.}

\Date{10/93}
%
%
%
%
%
\noblackbox
\def\ot{\otimes}
\def\zb{{\bar{z}}}

\def\zbar{{\bar{z}}}

%
%
%
%
%
%
%
%
%
%

\newsec{Introduction}

The massive integrable quantum field theories in 2 dimensions
are characterized as possessing an infinite number of local,
commuting integrals of motion, which for instance imply the
factorizability of the multi-particle S-matrix\ref\rzz{\ZZ}.
Over the past few years it has been shown that these theories
also possess an infinite number of non-abelian symmetries,
which generally correspond to $q$-deformations of affine
Lie algebras\ref\rbl{
\BLnlc}\ref\rfl{G. Felder and A. LeClair,
Int. Journ. Mod. Phys. A7 Suppl.
1A (1992) 239.}\ref\rsmir{
F. A.
Smirnov, Int. J. Mod. Phys.
A7, Suppl. (1992) .}.
It is now understood for example that the S-matrices are the
minimal solutions to the quantum affine symmetry equations.

In order to make further progress toward understanding the
full implications of these quantum affine symmetries, the
author investigated the following problem\ref\rlec{A. LeClair,
{\it Spectrum Generating Affine Lie Algebras in Massive
Quantum Field Theory}, CLNS 93/1220, hep-th/9305110}.
Consider the sine-Gordon (SG) theory, with the action
\eqn\eIi{
S = \inv{4\pi} \int d^2 z \(  \d_z \phi \d_\zb \phi
                 + 4\la  \cos ( \betah \phi ) \) . }
In \rbl\rfl\
explicit conserved currents were constructed for the
$6$  generators corresponding to the simple roots of
the $q-\hat{sl(2)}$ affine Lie algebra, where
$q = \exp ( -2\pi i / \betah^2 )$.  In this realization
the central extension, or level,  is zero, and the symmetry is
actually a deformed loop algebra.
When $\betah = 1$, the SG theory is equivalent to a free massive
Dirac fermion\ref\rcol{\Colemani}.  At this value of the
SG coupling constant, $q$ becomes $1$, and the results of \rbl\
predict the existence of an ordinary undeformed affine
$\hat{sl(2)}$ symmetry in the free Dirac theory.
By studying this limit we were able to develop some new structures
that we believe can be extended away from $q=1$.  Furthermore,
this is not a completely trivial exercise, since there exist
fields in the SG description,  such as $\exp( i\al \phi )$
for $\al \notin \Zmath$, which are not simply expressed in
terms of the free fermion fields and thus do not have
free-field form-factors or correlation functions.

In this talk we will mainly outline the  results in \rlec.
We will first  describe  the full infinite set of conserved $\slh$ charges
directly in the free massive
Dirac theory, and also the usual  infinite number of
abelian conserved charges $P_n$.  Radial quantization
will then be  introduced as the natural way to obtain operators which
diagonalize the Lorentz boost operator $L$.
This introduces a fermionic fock space
description of the space of fields $\hf$.  Furthermore, $\hf$ factorizes into
$\hf^L \ot \hf^R$, where in the massless limit
$\hf^L$ ($\hf^R$) is the left (right) `moving' space of field-states.
The conserved charges also factorize in
their action on $\hf^L \ot \hf^R$.  This leads to two separate
algebras $\hat{sl(2)}_L$ and $\hat{sl(2)}_R$, which each have level $1$ in
the anti-periodic sector, and level $0$ in the periodic sector.  In the
same fashion, the integrals of motion $P_n^{L,R}$ satisfy an infinite
Heisenberg algebra in the anti-periodic sector, whereas they all continue
to commute in
the periodic sector.  The spectrum of $\hf$ in the anti-periodic sector
can be obtained by supplementing the $\slh$ algebra with this
Heisenberg algebra, the fields being organized
into infinite highest weight modules.
By introducing a particle-field duality, we describe
a new way to compute form-factors in radial quantization.
We then  use these algebraic structures
to formulate an exact momentum space
bosonization.  In this operator formulation, non-trivial  form-factors
of the SG fields $\exp (\pm i \phi /2 )$
are computed as expectation values of vertex operators between
level 1 highest weight states.

\bigskip
\newsec{Affine $\hat{sl(2)}$ Symmetry of the Massive Dirac Fermion}

The Dirac theory is a massive free field theory of charged fermions.
Introducing the Dirac spinors
$ \Psi_\pm = \left(\matrix{\psib_\pm \cr \psi_\pm \cr}\right)$
of $U(1)$ charge $\pm 1$,
the action reads
\eqn\eIIi{
S = - \inv{4\pi} \int dx dt \(
\psibm \d_z \psibp + \psim \d_\zb \psip
+ i \ma ( \psim \psibp - \psibm \psip ) \) . }
We have continued to Euclidean space $t \to -it$, and $z,\zb$ are the
usual Euclidean light-cone coordinates:
$z = (t + ix)/2 , ~~~~~\zb = (t-ix)/2 $.

Generally, the conserved quantities follow from conserved currents
$J_\mu$:
\eqn\eIIiv{
\d_\zb J_z  + \d_z J_\zb = 0. }
We will often display    the two components
of conserved currents by writing
\eqn\ecurrent{
Q = \int \dz ~  J_z ~ - ~ \int \dzb ~ J_\zb , }
i.e. without specifying the contour of integration.

Using the
equations of motion:
\eqn\eIIxi{
\d_z \psib_\pm = i \ma \psi_\pm ,~~~~~
\d_\zb \psi_\pm = -i \ma \psib_\pm , }
it is easy to find an infinite number of conserved quantities.
They are the following:
\eqn\eIIxii{\eqalign{
Q^\pm_{-n} &= \frac{(-1)^{n+1}}{2}
\( \int \dz ( \psi_\pm \d_z^n \psipm ) - \int \dzb ( i\ma \> \psibpm \d_z^{n-1}
\psipm ) \)
\cr
Q^\pm_n &= \frac{(-1)^{n+1}}{2}
\( \int \dz ( -i\ma \> \psipm \d_\zb^{n-1} \psibpm )
- \int \dzb ( \psibpm \d_\zb^n \psibpm ) \)
\cr
\al_{-n} &= (-)^n \(
\int  \dz ( \psip \d_z^n \psim ) - \int \dzb ( i\ma \>\psibp \d_z^{n-1} \psim )
\)
\cr
\al_n &= (-)^n \(
\int \dz ( -i\ma \>  \psip \d_\zb^{n-1} \psibm )
- \int \dzb ( \psibp \d_\zb^n \psibm ) \)
, \cr } }
where $n\geq 0$, and $n$ is odd for $Q^\pm_n$.  (It turns out that
$Q^\pm_n = 0$ for $n$ even.)
The expressions for $Q^\pm_{\pm 1}$ were originally derived by
fermionizing the SG construction in \rbl\ at $\betah = 1$ and
noticing that the conservation of the resulting currents was
a simple consequence of Dirac equations of motion\ref\rraja{R. K.
Kaul and R. Rajaraman, Int. J. Mod. Phys. A8 (1993) 1815.}.
Similar conserved charges were constructed for $O(N)$ invariant
fermions in \ref\raass{\AASS}.

Define
\eqn\eIIxxii{
P_n \equiv \al_n  ~~~~~ n ~ {\rm odd} , ~~~~~
T_n \equiv \al_n  ~~~~~~~~n ~ {\rm even} . }
Then one can show using the fermionic commutation relations
that the charges satisfy the following  algebraic relations
\eqn\eIIxxiii{\eqalign{
\[ P_n , P_m \] &=
\[ P_n , T_m \] = \[ P_n , Q^\pm_m \] =
\[ T_n , T_m \] = 0  \cr
\[ T_n , Q^\pm_m \] &= \pm 2 ~ \mn ~ Q^\pm_{n+m}  \cr
\[ Q^+_n , Q^-_m \] &= \mn ~ T_{n+m}  \cr
\[ L , \al_n \] &= -n \> \al_n , ~~~~~\[ L , Q^\pm_n \] = -n \> Q^\pm_n
, \cr }}
where $L$ is the Lorentz boost operator.

We now interpret this algebraic structure. The $P_n$'s are the usual infinity
of commuting integrals of motion with Lorentz spin equal to an odd integer,
where $P_z = P_{-1}, ~ P_\zb = P_1 $, and the hamiltonian is $P_1 + P_{-1}$.
The additional charges $T_n , Q^\pm_n $
all commute with the $P_m$'s; for $m= \pm 1$ this is just the statement
that they are all conserved.
The charge $T_0$ is simply the $U(1)$ charge.
The commutation
relations of the $T_n, \Qpm n$ are the defining relations of the level
$0$ $\slh$ affine Lie algebra. More precisely this is a
twisted $\slh$ algebra.
The twist has a simple explanation:  the usual untwisted $\slh$ algebra
has an $sl(2)$ subalgebra of Lorentz scalars, whereas the Dirac theory
has only a $U(1)$ symmetry;  the twist breaks $sl(2)$ to $U(1)$.
It was shown in \rlec\ how Ward identities for the $\slh$ symmetry
can fix the correlation functions of the fermion fields.

\newsec{Radial Quantization}

Let $\CF$ denote the complete space of fields, and let $\CH_F$ be
spanned by the action of fields on the vacuum:
\eqn\ehf{
\CH_F = \{ \Phi_i (0) \rvac \equiv
|\Phi_i \rangle  , ~~~~ \Phi_i \in \CF \} . }
The space
$\CH_F$ diagonalizes the Lorentz boost operator $L$, but does
not diagonalize the momentum operators $P_z , P_\zbar$.
In order to construct the space $\CH_F$ explicitly, one can
work in radial quantization, since such a construction yields
states which diagonalize $L$.  Radial quantization was
generally considered in \ref\rfhj{S. Fubini, A. J. Hanson, and
R. Jackiw, Phys. Rev. D7 (1973) 1732.}, and specifically in
2d free fermion theories in \ref\rti{\TI}\ref\rgrif{\Griffin}.
 Define
the radial coordinates $(r, \vphi )$ as follows:
$z = \frac{r}{2} ~ \exp({i\vphi}) , ~~~~~\zb =
\frac{r}{2} ~ \exp({-i \vphi })$.
In radial quantization one treats the $r$-coordinate as a `time', and
$\vphi$ as the `space', and canonical commutation relations
are specified at equal $\vphi$.

One begins by expanding the fermion fields in a basis of
solutions to the Dirac equation of motion in radial coordinates.
One finds that one can define both a periodic (p) and an
anti-periodic (a) sector  in this way. Namely,
\eqn\eIIIxvii{
\Psi^{(a,p)}_\pm = \col{\psibpm}{\psipm} = \sum_\om
\bpm_\om ~ \Psi^{(a,p)}_{-\om - 1/2} ~+~
\bbpm_\om ~ \bar{\Psi}^{(a,p)}_{-\om -1/2} , }
where for the periodic sector $\om \in \Zmath + 1/2$, and
for the anti-periodic sector $\om \in \Zmath$.
The basis spinors have the following explicit expressions:
\eqn\eIIIxviii{\eqalign{
\Psi^{(a)}_{-\om -1/2} &=
\Gamma (\ha - \om ) ~ \ma^{\om + 1/2} ~
\col{i \> e^{i(\ha - \om )\vphi} ~ I_{\ha - \om} (\ma r) }
{e^{-i(\om + \ha )\vphi} ~ I_{-\om -\ha} (\ma r)}  \cr
\bar{\Psi}^{(a)}_{-\om - 1/2}
&=
\Gamma (\ha - \om ) ~ \ma^{\om + 1/2} ~
\col{ e^{i(\ha + \om )\vphi} ~ I_{-\ha - \om} (\ma r) }
{-i \> e^{-i(\ha - \om )\vphi} ~ I_{\ha - \om } (\ma r)}  ,\cr}}
\eqn\eIIIxix{\eqalign{
\cr
\Psi^{(p)}_{-\om -1/2} &= \frac{2 \ma^{\om + 1/2}}{\Ga (\ha + \om )}
{}~ \col{-i \> e^{i(\ha - \om )\vphi} ~ K_{ \om - \ha} (\ma r) }
{e^{-i(\om + \ha)\vphi} ~ K_{\om +\ha} (\ma r)} ~~~~\om \geq 1/2   \cr
\bar{\Psi}^{(p)}_{-\om -1/2} &= \frac{2 \ma^{\om + 1/2}}{\Ga (\ha + \om )}
{}~ \col{ e^{i(\ha + \om )\vphi} ~ K_{ \om + \ha} (\ma r) }
{i \> e^{i(\om - \ha)\vphi} ~ K_{\om -\ha} (\ma r)} ~~~~\om \geq 1/2
, \cr}}
and $\Psi^{(p)}_{-\om - 1/2}$, $\bar{\Psi}^{(p)}_{-\om -1/2} $
for $\om \leq -1/2$ have the same expression as in the anti-periodic sector.
In the quantum theory these modes satisfy:
\eqn\eIIIxxiv{
\{ b^+_\om , b^-_{\om'} \} =
\{ \bbp_\om , \bbm_{\om'} \} = \de_{\om , -\om'}
, ~~~~~\{ b_\om , \bb_{\om '} \} = 0. }

In the massless limit $\ma \to 0$, the operators $b^\pm_\om , \bb^\pm_\om$
are the familiar fermionic oscillators of conformal field theory.
Indeed,
\eqn\eIIIxxx{
\col{\psibpm}{\psipm} ~ \mtoo ~ \sum_\om ~
\col{\bbpm_\om ~ \zb^{-\om-1/2} }{\bpm_\om ~ z^{-\om-1/2} } . }
What is remarkable is that since the oscillators
are defined in the massive theory, the fermion fock spaces
built from these oscillators continue to correspond
precisely to the space of fields $\CH_F$ in the
{\it massive} theory.  Furthermore, $\CH_F$ factorizes
into quasi-left/right pieces:
\eqn\equasi{
\CH_F = \CH^L_F \ot \CH^R_F . }
For more discussion on this point, see \rlec.

Consider first the periodic sector.
We define the physical vacuum as follows:
\eqn\eIIIxxv{
\bpm_\om \> \rvac = \bbpm_\om \> \rvac = 0 , ~~~~~\om \geq 1/2 . }
Define the left and right periodic fock spaces:
\eqn\eIIIxxvi{\eqalign{
\CH^L_p &= \left\{ \bm_{-\om_1}
\bm_{-\om_2} \cdots \bp_{-\om_1 '} \bp_{-\om_2 '}
\cdots \rvac \right\} \cr
\CH^R_p &= \left\{ \bbm_{-\om_1}
\bbm_{-\om_2} \cdots \bbp_{-\om_1 '} \bbp_{-\om_2 '}
\cdots \rvac \right\} , \cr}}
where $\om , \om'  \geq 1/2$.
One can see directly that in the periodic sector
$\CH_F = \CH^L_p \ot \CH^R_p$.  For example, using the
explicit expansions \eIIIxvii, one finds
\eqn\eIIIxxxi{
\d_z^n \psipm (0) \rvac = n! ~ \bpm_{-n-\ha} \rvac , ~~~~~
\d_\zb^n \psibpm (0) \rvac = n! ~  \bbpm_{-n-\ha} \rvac . }
Other states in $\CH^{L,R}_p$ correspond to composite operators.
For example, consider the $\uone$ current $J_z = \psip \psim $,
$J_\zb = \psibp \psibm $.  One has
\eqn\eIIIxxxiii{
J_z (0) \rvac = \bp_{-\ha} \> \bm_{-\ha} \> \rvac ,
{}~~~~~J_\zb (0) \rvac = \bbp_{-\ha} \> \bbm_{-\ha} \> \rvac . }

We now turn to the anti-periodic sector.  Due to the existence of the
zero modes $\bpm_0 , \bbpm_0$, the `vacuum'
in this sector is doubly
degenerate for both left and right.  Define these vacua as
$\rvacpm_L$ and $\rvacpm_R$, characterized by
\eqn\eIIIxxxiv{\eqalign{
\bpm_0 \> \va{\mp \ha}_L = \rvacpm_L , ~~~~ \bpm_0 \> \rvacpm_L = 0 ,
{}~~~~ \bpm_n \> \rvacpm_L = 0,~~~~&n\geq 1 \cr
\bbpm_0 \> \va{\mp \ha}_R = \rvacpm_R , ~~~~ \bbpm_0 \> \rvacpm_R = 0 ,
{}~~~~ \bbpm_n \> \rvacpm_R = 0,~~~~&n\geq 1 . \cr}}
The vacua $\lvacpm$ are defined by the inner products
\eqn\eIIIxxxv{
{}_L \lva{\mp \ha} \pm \ha \rangle_L = {}_R \langle \mp \ha \rvacpm_R = 1.}
These vacuum states have $U(1)$ charge $\pm 1/2$.  The anti-periodic
fock spaces are defined as
\eqn\eIIIxxxvb{\eqalign{
\CH^L_{a_{\pm}}
&= \left\{ \bm_{-n_1} \bm_{-n_2} \cdots \bp_{-n_1 '} \bp_{-n_2 '}
\cdots \rvacpm_L \right\} \cr
\CH^R_{a_{\pm}}
&= \left\{ \bbm_{-n_1} \bbm_{-n_2} \cdots \bbp_{-n_1 '} \bbp_{-n_2 '}
\cdots \rvacpm_R \right\} , \cr}}
for $n , n' \geq 1$.
Based on a study of the massless limit the following identification
was proposed in \rlec:
\eqn\eIIIxxxix{
e^{\pm i \phi (0)/2 } \> \rvac ~=~ \( \rvacpm_L \ot \rvacmp_R \)
\equiv \rvacpm , }
where $\phi (z, \zb )$ is the local SG field.

Consider now the conserved charges of the last section in
radial quantization.
Given
a conserved current $J_z , ~ J_\zb$, the conserved charge is
\eqn\eVi{
Q = \inv{4\pi} \int_{-\pi}^\pi ~ r \> d\vphi
\( e^{i\vphi} \>  J_z  ~+~ e^{-i\vphi} \>  J_\zb \) . }
All of the conserved charges constructed in section 2 can thereby be
expressed in terms of the radial modes $b^\pm_\om , ~ \bb^\pm_\om$.
More specifically, define the inner product of two spinors
$A = \col{\bar{a}}{a}$, $B = \col{\bar{b}}{b}$ as
\eqn\eVii{
(A,B) = \inv{4\pi} \int_{-\pi}^\pi ~ r d\vphi
\( e^{ i\vphi} ~ a \>  b ~+~ e^{-i\vphi} ~ \bar{a} \> \bar{b} \) . }
The conserved charges can all be expressed using the above inner
product:
\eqn\eVv{\eqalign{
Q^\pm_{-n} &= \frac{(-)^{n+1}}{2}  \( \Psi_\pm , \d^n_z \Psi_\pm \) , ~~~~~
Q^\pm_{n} = \frac{(-)^{n+1}}{2} \( \Psi_\pm , \d^n_\zb \Psi_\pm \) \cr
\al_{-n} &= {(-)^{n}}  \( \Psi_+ , \d^n_z \Psi_- \) , ~~~~~
\al_{n} = {(-)^{n}} :\( \Psi_+ , \d^n_\zb \Psi_- \) : ~
. \cr }}
One finds that the charges split into left and right pieces:
\eqn\eVvii{
Q^\pm_n ~=~ Q^{\pm , L}_n ~+~ Q^{\pm ,R}_{-n}  ,~~~~~~~
\al_n ~=~ \al^L_n ~+~ \al^R_{-n} .}
The additional minus sign in the subscript $-n$ of the right piece
of the charges in comparison to the left piece is dictated by Lorentz
covariance.
The explicit expressions are as follows:
\eqn\eVviii{\eqalign{
\al^L_n &=
 {\ma^{|n| + n}} ~ \sum_{\om \in S^{(a,p)}_n }
\frac{\Ga (\ha + \om -n )}{\Ga (\ha + \om )} ~
 : b^+_{n-\om} ~ b^-_{\om } : \cr
Q^{\pm , L}_n &=
 {\ma^{|n| + n}} ~ \sum_{\om \in S^{(a,p)}_n }
\frac{\Ga (\ha + \om -n )}{\Ga (\ha + \om )} ~
  b^\pm_{n-\om} ~ b^\pm_{\om }  , \cr }}
  where the sums over $\om$ run over $S^{(p)}_n$ ($S^{(a)}_n $) for
  the periodic (anti-periodic) sector, and
\eqn\esn{\eqalign{
S^{(a)}_n &= \Zmath , ~~~\forall ~ n \cr
S^{(p)}_n &= \{ \om \in \Zmath + 1/2 ; ~~0>\om >n ~ {\rm if}~ n> 0 ;
{}~~n>\om >0 ~ {\rm if} ~ n< 0 \} . \cr }}
Identical expressions apply to $Q^{\pm , R}_n , \al^{R}_n $ with
$b^\pm_\om \to \bb^\pm_\om $.

Define as before
$P_n^{L,R} = \al^{L,R}_n , ~~n ~~{\rm odd}; ~~~
T_n^{L,R} = \al^{L,R}_n , ~~n ~~{\rm even}$.
Then one can show that the above operators  satisfy the relations
\eqn\eVxvii{\eqalign{
\[ P^L_n , P^L_m \] &= \[ T^L_n , T^L_m \] =
n ~ k ~ \ma^{2|n|} ~ \de_{n, -m}  \cr
\[ P^L_n , T^L_m \] &= \[ P^L_n , Q^{\pm,L}_m \] = 0  \cr
\[ T^L_n , Q^{\pm ,L}_m \] &= \pm 2 ~ \mn ~ Q^{\pm, L}_{n+m}   \cr
\[ Q^{+,L}_n , Q^{-,L}_m \] &= \mn ~ T^L_{n+m} ~+~
\frac{n}{2} ~k~  \ma^{2|n|} ~ \de_{n,-m}  , \cr}}
where in the periodic sector the level $k$ is zero, and in the
anti-periodic sector $k=1$.  Identical results apply to
the right pieces of the conserved charges, with the same level
1 in the antiperiodic sector.  Note that the sum of the left and
right operators \eVvii\ continues to satisfy a level $k=0$ algebra
in either sector, which is required for consistency with section
2.

Due to the non-zero level 1 in the anti-periodic sector, the space
of fields in this sector can be organized using infinite highest
weight modules.  At level 1 there are only
 2 highest weight
states $\va{\La_j}$, $j = 0,1/2$, satisfying
\eqn\eVxx{\eqalign{
Q^{\pm, L}_n \> \va{\La_j}_L &= T^L_n \> \va{\La_j}_L =0
{}~~~~~n\geq 1  \cr
T^L_0 \> \va{\La_0}_L = -\ha ~ \va{\La_0}_L ,~~~~&~~~~
T^L_0 \> \va{\La_\ha}_L = \ha ~ \va{\La_\ha}_L . \cr}}
These highest weight states are the `vacuum' states defined
above with the identification:
\eqn\eVxxi{
\rvacp_L = \va{\La_\ha} , ~~~~~\rvacm_L = \va{\La_0} . }

Let us denote the $P_n$ extension of $\slh$ as $\slhh$, and
define
the $\slhh_L$ modules $\vhh^{{}_L}_{\La_j}$ as follows,
\eqn\eVxxvi{
\vhh^{{}_L}_{\La_i} \equiv \left\{ Q^{\pm,L}_{-n_1} ~Q^{\pm , L}_{-n_2} \cdots
T^L_{-n_1 '} T^L_{-n_2 '} \cdots P^L_{-n_1 ''} P^L_{-n_2 ''} \cdots
| \Lambda_i \rangle  \right\} , }
for $n,n' , n'' \geq 1$.
Then one can show that
\eqn\eident{
\hal = \vhh^{{}_L}_{\Lambda_0} \oplus \vhh^{{}_L}_{\Lambda_\ha} , }
where $\hal = \CH^L_{a_+} \oplus \CH^L_{a_-}$.
This is proven by comparing the ${\rm Tr} ~(q^L )$ characters for the
fermionic space $\hal$ with the ones for the (twisted) $\slhh$ modules.

\newsec{Particle-Field Duality and Form-Factors in Radial Quantization}

In conventional canonical approaches to massive quantum field
theory, one deals with the space of particles $\CH_P$.  In
2d, parameterizing the energy and momentum in terms of the
usual rapidity, $E= m\cosh \th$, $P= m \sinh \th$, the
space $\CH_P$ is spanned by
\eqn\duali{
\CH_P = \{ |\th_1 , \cdots , \th_n \rangle_{\ep_1 \cdots \ep_n } \},
}
where $\ep_i$ are the quantum numbers of the particles.
One can easily define a dual space $\CH^*_P$, and a completeness
relation
\eqn\edualiii{
1 = \sum_{n=0}^\infty \inv{n!} \sum_{\ep_i} \int d\th_1 \cdots
d\th_n ~ |\th_1 ,\cdots , \th_n \rangle_{\ep_1 \cdots \ep_n}
{}^{\ep_1 \cdots \ep_n } \langle \th_1 , \cdots , \th_n | . }

Form-factors are matrix elements of fields in the space $\CH_P$.
Consider for example the form-factor
\eqn\edualii{
{}^{\ep_1 \cdots \ep_n } \langle \th_1 \cdots \th_n | \Phi (0) \rvac ,}
where
$\Phi \in \CF$.  The completeness relation in $\CH_P$ allows one
to view $\Phi (0) \rvac$ as a element of $\CH_P$.  Conventionally,
one thinks of expressing the fields $\Phi (0)$ in terms of the operators
that create the states in $\CH_P$, and one computes form-factors in
the space $\CH_P$.

We now propose an entirely different way to compute the same form-factors.
Let us suppose that one can define a non-degenerate inner-product on
the space of fields $\CH_F$, and thereby construct the dual space
$\CH^*_F$ and the completeness relation:
\eqn\edualiiib{
1 = \sum_i |\Phi_i \rangle \langle \Phi^i | . }
This implies that one can then consider states in $\CH^*_P$ as
elements of $\CH_F^*$:
\eqn\edualiv{
{}^{\ep_1 \cdots \ep_n } \langle \th_1 \cdots \th_n | ~\in
{}~ \CH^*_F . }
Consequently, the form-factors can be computed directly in the space
$\CH_F$ rather than $\CH_P$.  Radial quantization provides us
with an explicit construction of the spaces $\CH_F , \CH^*_F$ and
the completeness relation, thus by the above reasoning,
radial quantization leads to a new way to compute form-factors.
The success of the method relies on being able to construct
explicitly the map \edualiv. We now describe how to determine this
map in the model we are considering.  Henceforth we will be
working exclusively in the antiperiodic sector\foot{The
periodic sector will be considered in \ref\rcostas{C. Efthimiou
and A. LeClair, {\it Particle-Field Duality and Form-Factors
from Vertex Operators}, in preparation.}}.

In the standard temporal quantization, the fermion fields have
the following plane wave expansions
\eqn\eIIxx{
\Psi_\pm = \col{\psibpm}{\psipm} = \pm \sqm
\int_{-\infty}^\infty \du \> \bh^\pm (u) ~
\col{1/ \squ}{-i\squ} ~ \ez , }
where $u= \exp (\th )$.  We have combined the usual
creation-annihilation operators into a single operator
$\bh^\pm (u)$, where $\bh^\pm (u>0)$ is a creation
operator and $\bh^\pm (u<0)$ is an annihilation operator.
(See \rlec\ for details.) The  states in
$\CH^*_P$ are constructed as follows:
\eqn\edualv{
{}^{\ep_1 \cdots \ep_n } \langle \th_1 \cdots \th_n |
= \inv{(2\pi i )^n } ~
\lvac \bh^{\ep_1} (-u_1 ) \cdots \bh^{\ep_n} (-u_n ) , }
where $u_i = \exp( \th_i )$.

The key to constructing explicitly the map from
$\CH^*_P$ to $\CH^*_F$ is based on the following fact.
By analytically continuing the $u$ integral in \eIIxx\
and appropriately redefining $\bh^\pm (u)$, one can
obtain the expansions in radial quantization. More
specifically, let
\eqn\eIIIxiii{
\int_{-\infty}^\infty \du ~ \bhpm (u)
\rightarrow
\(
\int_{\cL} \dua ~ b_\pm (u) ~~ - ~~
\int_{\cR} \dua ~ \bb_\pm (u) \), }
where
$\cL , \cR$ are contours depending on the angular direction $\vphi$ of
the cut displayed in figures 1,2,  and
\midinsert
\epsfxsize = 2in
\vbox{\vskip -.1in\hbox{\centerline{\epsffile{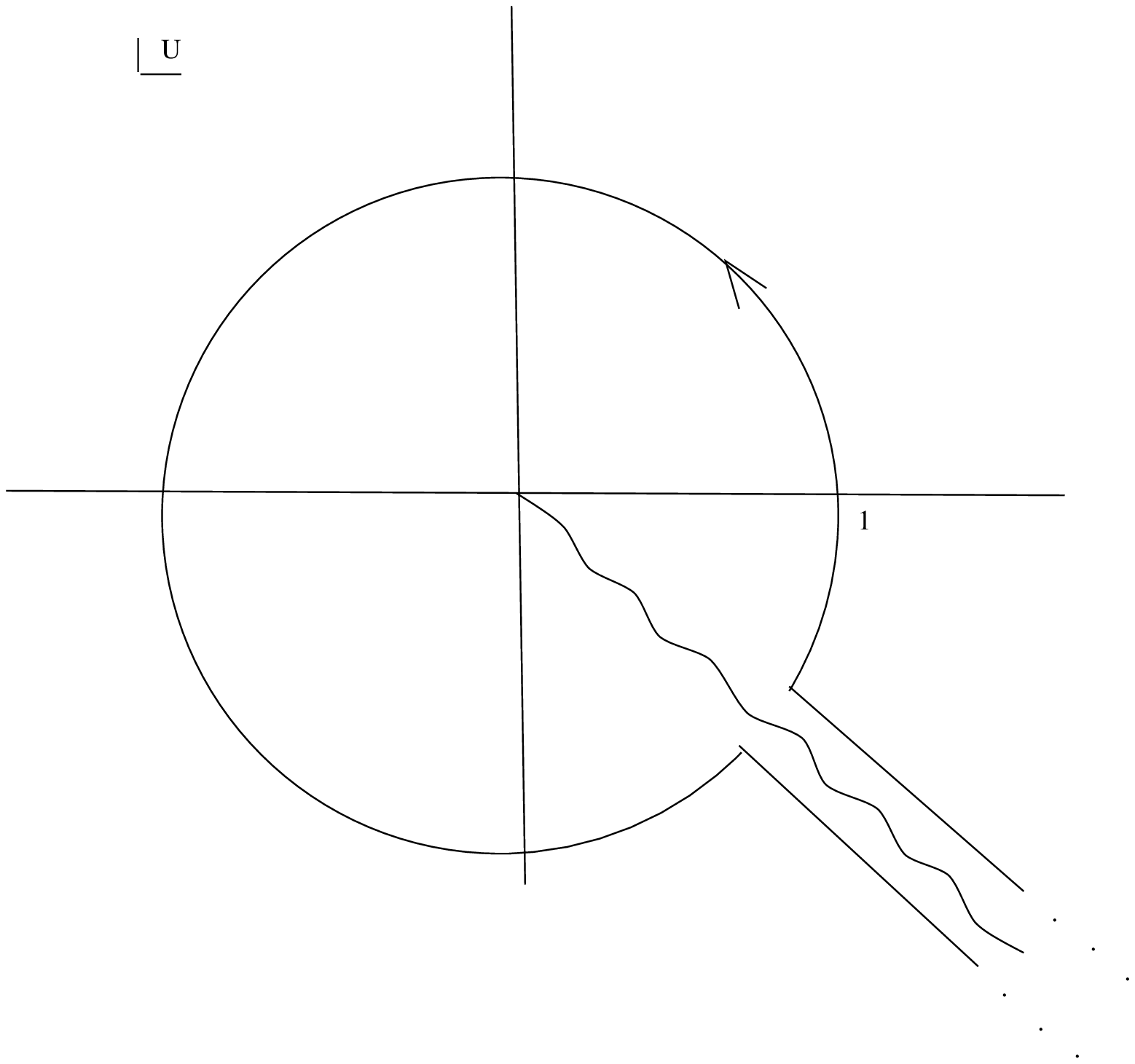}}}
\vskip .1in
{\leftskip .5in \rightskip .5in \noindent \ninerm \baselineskip=10pt
\bigskip\bigskip
Figure 1.
The contour $\CC^L_\vphi$.  The cut (wavy line) is oriented at an
angle $\vphi$ from the negative $y$-axis. The circle is at $|u| = 1$.
\smallskip}} \bigskip
\endinsert
\midinsert
\epsfxsize = 2in
\bigskip
\vbox{\vskip -.1in\hbox{\centerline{\epsffile{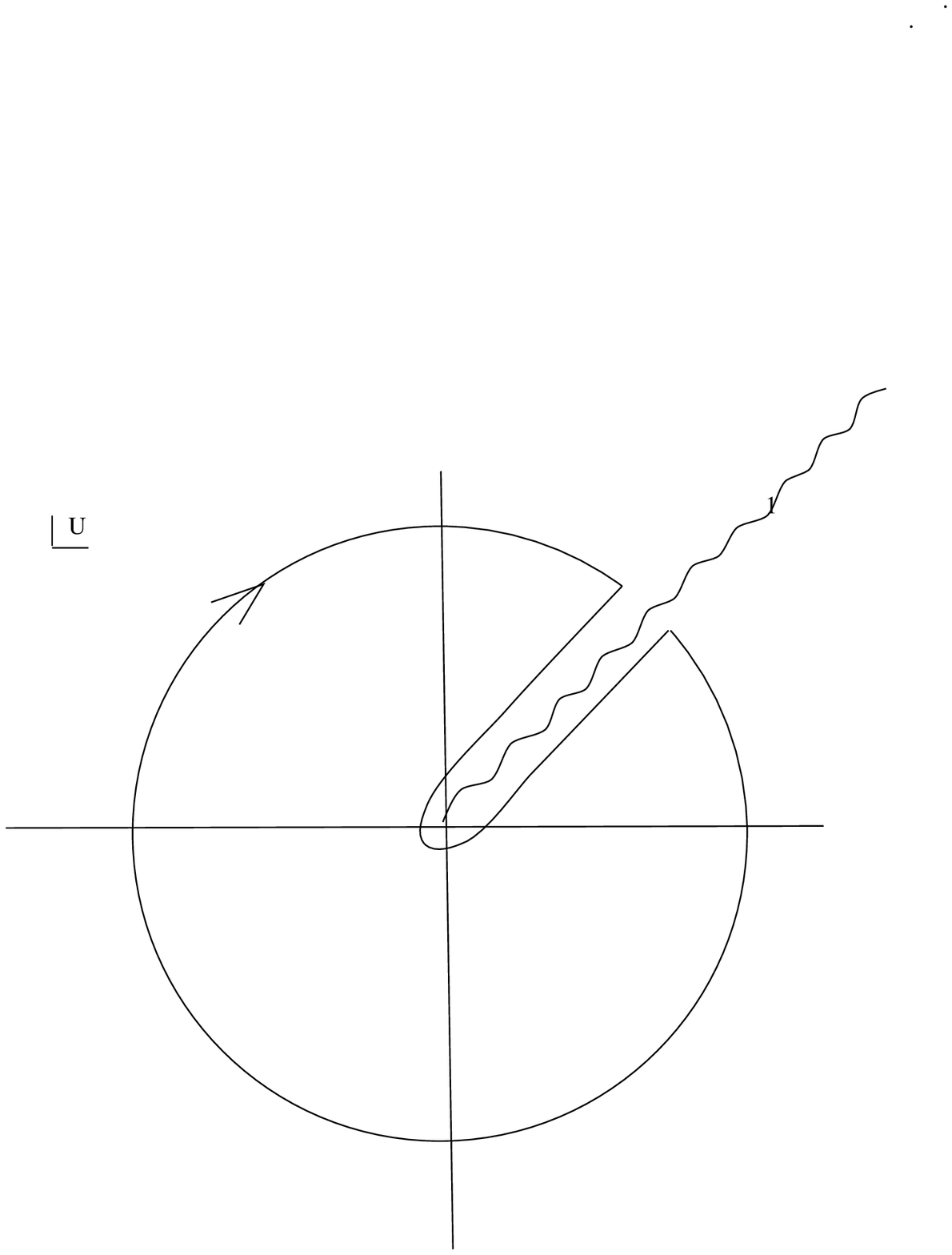}}}
\vskip .1in
{\leftskip .5in \rightskip .5in \noindent \ninerm \baselineskip=10pt
Figure 2.
The contour $\CC^R_\vphi$.  The cut  is oriented at an
angle $\vphi$ from the positive $y$-axis.  The circle is at
$|u| = 1$.
\smallskip}} \bigskip
\endinsert
\eqn\eIIIxiv{\eqalign{
\bpm (u) &= \pm i \sum_{\om \in \Zmath } \Gamma (\ha - \om )
\> \ma^\om ~ b_\om ~ u^\om   \cr
\bbpm (u) &= \pm  \sum_{\om \in \Zmath } \Gamma (\ha - \om )
\> \ma^\om ~ \bb_\om ~ u^{-\om}  .    \cr }}
Using integral representations for the Bessel functions, one
obtains \eIIIxviii.
This leads us to propose the following fundamental expression:
\eqn\edualvi{
\lvac \bh^{\ep_1} (u_1 ) \cdots \bh^{\ep_n} (u_n)
= \inv{2} \( \langle +\ha | + \langle -\ha | \)
\( b^{\ep_1} (u_1 ) + \bb^{\ep_1} (u_1 ) \)
\cdots
\( b^{\ep_n} (u_n ) + \bb^{\ep_n} (u_n ) \) . }
The RHS of the above equation is a state in $\CH^*_F$, and
this means that the computation of the form-factors can
be carried out in $\CH_F$.

As an example, consider the form-factors of the fields
$\exp (\pm i \phi /2 )$ in the SG theory.  From the
identification \eIIIxxxix\ one has
\eqn\edualvii{\eqalign{
{}^{\ep_1 \cdots \ep_n } \langle \th_1 \cdots \th_n |
{}~ \exp (\pm i \phi /2 ) \rvac
&= \inv{(2\pi i)^n } {}_L \langle \mp \ha |
b^{\ep_1} (-u_1) \cdots b^{\ep_n} (-u_n) | \pm \ha \rangle_L \cr
&= \inv{(2\pi i)^n } {}_R \langle \pm \ha |
\bb^{\ep_1} (-u_1) \cdots \bb^{\ep_n} (-u_n) | \mp \ha \rangle_R \cr
.\cr}}
Additional arguments for the above formula were given in \rlec.
One can compute the RHS explicitly and one finds agreement with
the known result\ref\rmss{\MSS}\ref\rform{\form}. We will
show this explicitly in the next section.  Note that
the RHS of the above expression has the structure of a conformal
free-fermion correlator, but here we are in momentum space!

\newsec{{\bf Vertex Operators and Momentum Space Bosonization}}

We will now describe a vertex operator construction for the
form-factors of the last section.  This involves constructing
a bosonic representation of the operators $b^\pm (u) , \bb^\pm (u)$.
Recall that in the anti-periodic section one has the infinite
Heisenberg algebras of the chirally split integrals of motion:
\eqn\eVxxxvi{
\[ \al^L_n , \al^L_m \] = n~ \ma^{2|n|} ~ \de_{n,-m} , ~~~~~
\[ \al^R_n , \al^R_m \] = n~ \ma^{2|n|} ~ \de_{n,-m} . }
Define some momentum space bosonic fields as follows:
\eqn\eVIx{\eqalign{
-i \phih^L (u) &=    \sum_{n\neq 0} \ma^{-|n|} ~ \al^L_n ~ \frac{u^n}{n}
  + \al_0^L \log (u) - \alt_0^L  \cr
-i \phih^R (u) &=    \sum_{n\neq 0} \ma^{-|n|} ~ \al^R_n ~ \frac{u^{-n}}{n}
  + \al_0^R \log (u) + \alt_0^R  , \cr }}
where
$\[ \al_0^L , \alt_0^L \] =
\[ \al_0^R , \alt_0^R \] = 1$.
The vacua $|\emptyset \rangle_{L,R}$ are defined to satisfy
\eqn\eVIxiii{
\al_n^L \emrvac_L = \al_n^R \emrvac_R = 0 , ~~n\geq 0; ~~~~~
\alt^L_0 \emrvac_L ~, ~~\alt^R_0 \emrvac_R \neq 0 . }
Note that $ \emrvac \equiv \emrvac_L \ot \emrvac_R $  does not
correspond to the physical vacuum since it is not annihilated by
$P_1$ and $P_{-1}$.
One has the following vacuum expectation values
\eqn\eVIxiv{\eqalign{
{}_L \emlvac ~ \phih^L (u) ~ \phih^L (u') ~\emrvac_L
&= -\log ( 1/u - 1/u' ) \cr
{}_R \emlvac ~ \phih^R (u) ~ \phih^R (u') ~\emrvac_R  &= -\log (u - u' ) \cr}}
\medskip
\eqn\eVIxv{\eqalign{
{}_L \emlvac \prod_i ~ e^{i\al_i \phih^L (u_i )} ~ \emrvac_L
&= \prod_{i< j} \( 1/u_i - 1/u_j \)^{\al_i \al_j }  \cr
{}_R \emlvac \prod_i ~ e^{i\al_i \phih^R (u_i )} ~ \emrvac_R
&= \prod_{i< j} \( u_i - u_j \)^{\al_i \al_j }  . \cr}}

Using these free momentum space fields one can bosonize all
of the ingredients of the last section:
\eqn\eVIxix{\eqalign{
b^+ (u) &=  \sqrt{\frac{\pi}{u} } : e^{i\phih^L (u)}: , ~~~~~
b^- (-u) = \sqrt{\frac{\pi}{u} } : e^{-i\phih^L (u)}: \cr
\bb^+ (u) &= \sqrt{\pi u } : e^{-i\phih^R (u)}: , ~~~~~
\bb^- (-u) = - \sqrt{\pi u} : e^{i\phih^R (u)}:  , \cr }}
and
\eqn\eVIxvi{\eqalign{
\va{\al}_L ~ &= ~ : \exp (i\al \phih^L  (\infty )): ~ \emrvac_L ,
{}~~~~~
\va{\al}_R ~ = ~: \exp (-i \al \phih^R (0 )): ~ \emrvac_R \cr
{}_L
\lva{\al} &= \lim_{u\to 0} ~ u^{-\al^2} ~ \emlvac ~: \exp ( i\al \phih^L (u)):
, ~~~~~
{}_R
\lva{\al} = \lim_{u\to \infty} ~ u^{\al^2} ~ \emlvac ~
\exp (-i \al \phih^R (u))
.\cr}}
 From this construction one finds
\eqn\effii{\eqalign{~
& {}_L \langle \mp \ha \vert
b^+ (u_1 ) \cdots b^+ (u_n ) b^- (u_{n+1} ) \cdots b^- (u_{2n} )
\vert \pm \ha \rangle_L \cr
&~~~~~~=
 {}_R \langle \pm \ha \vert
\bb^+ (u_1 ) \cdots \bb^+ (u_n ) \bb^- (u_{n+1} ) \cdots \bb^- (u_{2n} )
\vert \mp \ha \rangle_R  \cr
&~~~~~~=
\pi^n
\sqrt{u_1 \cdots u_{2n} }
\( \prod_{i=1}^n \( \frac{u_{i+n}}{u_i} \)^{\pm 1/2} \)
\( \prod_{i<j \leq n} (u_i - u_j ) \)  \cr
& ~~~~~~~~~~\cdot \( \prod_{n+1 \leq i < j } (u_i - u_j ) \)
\( \prod_{r=1}^n \prod_{s=n+1}^{2n} \inv{u_r + u_s } \)   . \cr } }
These expressions agree with the known form-factors,
though they were
originally computed using rather different methods\rmss\rform.

\newsec{Conclusions}

We have described a new way to compute form-factors based
on structures inherent in radial quantization, as
vacuum expectation values of vertex operators.
The integrals of motion played an essential role in
the explicit construction of the vertex operators.
We believe that these results can be extended away
from the free fermion point, eventually leading to a vertex
operator construction of the SG form-factors.  These
form-factors have been previously computed by Smirnov
using bootstrap methods\rform.
Recently, Lukyanov has proposed a vertex operator
construction for the SG form-factors at all values
of the SG coupling\ref\rluk{S. Lukyanov, {\it
Free Field Representation for Massive Integrable Models},
Rutgers preprint RU-93-30, hep-th/9307196.}.
Comparing the latter construction
with the results described here one finds some apparent
differences.  The free boson oscillators in \rluk\
have no apparent connection with the integrals of motion,
in contrast with the above results.  More importantly,
it is proposed in \rluk\ that the form-factors are
computed as traces over free-field modules, whereas in
our work they are vacuum expection values.  It would
be interesting to understand more precisely if and how the
construction in \rluk\ is related to ours at the free-fermion
point.

\bigskip
\centerline{Acknowledgements}

I would  like to thank the organizers of the Berkeley Strings 93
conference for the opportunity to present this work.
This work is supported
by an  Alfred P. Sloan Foundation fellowship,  and the
National Science Foundation in part through the
National Young Investigator program.

\listrefs
\bye